\documentclass[aps,prl,twocolumn,showpacs,superscriptaddress]{revtex4}

\usepackage{graphicx}
\usepackage{hyperref}

\begin{document}

\title{Compact Source of EPR Entanglement and Squeezing at Very Low Noise Frequencies}

\author{J. Laurat}\affiliation{Laboratoire
Kastler Brossel, UPMC, Case 74, 4 Place Jussieu, 75252 Paris cedex
05, France}
\author{T. Coudreau}\email{coudreau@spectro.jussieu.fr}\affiliation{Laboratoire
Kastler Brossel, UPMC, Case 74, 4 Place Jussieu, 75252 Paris cedex
05, France} \affiliation{Laboratoire Mat{\'e}riaux et
Ph{\'e}nom{\`e}nes Quantiques, Universit{\'e} Denis Diderot, 2
Place Jussieu, 75251 Paris cedex 05, France}
\author{G. Keller}\affiliation{Laboratoire
Kastler Brossel, UPMC, Case 74, 4 Place Jussieu, 75252 Paris cedex
05, France}
\author{N. Treps}\affiliation{Laboratoire
Kastler Brossel, UPMC, Case 74, 4 Place Jussieu, 75252 Paris cedex
05, France}
\author{C. Fabre } \affiliation{Laboratoire
Kastler Brossel, UPMC, Case 74, 4 Place Jussieu, 75252 Paris cedex
05, France}

\begin{abstract}
We report on the experimental demonstration of strong quadrature
EPR entanglement and squeezing at very low noise sideband
frequencies produced by a single type-II, self-phase-locked,
frequency degenerate optical parametric oscillator below
threshold. The generated two-mode squeezed vacuum state is
preserved for noise frequencies as low as 50 kHz. Designing simple
setups able to generate non-classical states of light in the kHz
regime is a key challenge for high sensitivity detection of
ultra-weak physical effects such as gravitational wave or small
beam displacement.
\end{abstract}


\maketitle

Since the pioneering work of Caves \cite{caves} which showed that
it is possible to improve the sensitivity of interferometric
measurements by the use of squeezed light, and its experimental
demonstrations \cite{grangier}, various protocols involving
squeezed light have been discussed in order to beat the standard
quantum limit in gravitational wave detectors \cite{kimble}. As
next generations of gravitational wave detectors will be designed
to be shot noise limited in the acoustic band from 10 Hz to 10
kHz, such techniques appear as quite promising ways to improve
their sensitivity. Recently, a squeezing-enhanced power-recycled
Michelson interferometer has been experimentally demonstrated and
signal-to-noise ratio improvement obtained \cite{mckenzie}.
However, the injected squeezing bandwidth lies around 5 MHz and
not in the frequency band of gravitational waves. A source of low
frequency squeezing is thus a key point for the implementation of
future squeezed-input interferometers.

More generally, many high sensitivity measurements performed at
low modulation frequency can benefit from such a device. In
\cite{treps}, a "quantum laser pointer" has been experimentally
demonstrated and an improvement of modulated small displacement
measurements in two orthogonal directions in the transverse plane
has been reported. Improved beam positioning sensitivity below the
shot noise limit is obtained at the frequency where squeezing is
available, a few MHz in the case of this experiment. Two sources
of low frequency squeezing are needed to apply this promising
displacement measurement technique to actual instruments where
frequency modulation is generally low. This could be applied for
instance to AFM microscopy in tapping mode where cantilever
oscillates at its resonant frequency which is typically a few
hundreds of kHz. Such a squeezing source should also improve the
thermo-optical spectroscopy technique called "mirage effect"
\cite{boccara}, which enables the measurement of very weak
absorption: a thermo-optical modulation on a sample induces a
periodic refractive index gradient and results on a low frequency
modulated probe beam deflection.

Broadband and low frequency squeezing is also very useful even if
the information is not carried by a single frequency modulation.
When a beam is detected during a finite time, the signal-to-noise
ratio depends on the noise in an extended range of sideband
frequencies. Broadband squeezing with a cut-off frequency as low
as possible is thus required. A great number of measurements can
be improved in that way, for instance the detection of weak pulsed
signals or the reduction  of the bit error rate in the readout of
digitized optical information.

\begin{figure}[htpb!]
\includegraphics[width=.95\columnwidth]{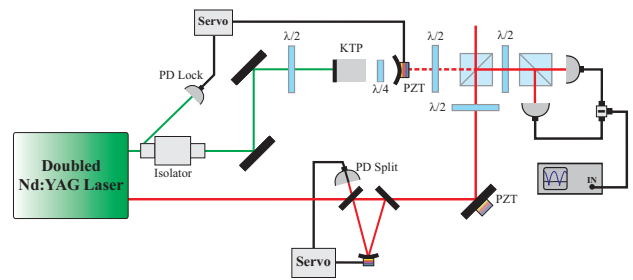}
\caption{\label{setup}Experimental layout. A continuous
frequency-doubled Nd:YAG laser pumps below threshold a frequency
degenerate type-II phase-matched OPO with a quarter-wave plate
inserted inside the cavity. The generated two-mode state is
characterized by homodyne detection and the difference of the
photocurrents is sent onto a spectrum analyser. PD Lock: FND-100
photodiode for locking of the OPO. PD Split: split two-element
InGaAs photodiode for tilt-locking of the filtering cavity.}
\end{figure}

In \cite{pfister}, quantum noise reduction on the intensity
difference of twin beams has been observed down to 90 kHz.
However, only two experiments to date have demonstrated
continuous-wave squeezing at low frequencies. Both experiments are
based on an optical noise cancellation scheme where sources of
squeezing are inserted within a Mach-Zehnder interferometer:
squeezing has been obtained around 220 kHz with a pair of
independent optical parametric amplifiers (OPA) \cite{bowen} and
very recently around 100 kHz with a single OPA \cite{schnabel}.
This frequency range is rather unusual in the experimental quantum
optics field where non-classical properties are generally observed
in the MHz range due to the presence of large classical noise at
lower frequencies. In this letter, we report on what is to our
knowledge the first experimental demonstration of very low
frequency continuous-wave squeezing without need of optical noise
cancellation scheme. Broadband vacuum squeezing is observed for
frequencies down to 50 kHz. Furthermore, our setup generates not
only single-mode squeezing but also EPR entanglement which is a
basic requisite in quantum information protocols such as
teleportation, dense-coding or optical-atomic interfacing
\cite{CV}.

Our setup relies on a frequency degenerate type-II phase-matched
optical parametric oscillator (OPO) below threshold, in which a
quarter-wave plate inserted in the cavity adds a linear coupling
between the signal and idler fields. In this letter, we focus on
the case where the plate is rotated by a very small angle (smaller
than $0.02^{\circ}$) relative to the principal axis of the
non-linear crystal. In such a configuration, the signal and idler
modes are entangled: they show quantum correlations and
anti-correlations on orthogonal quadratures for noise frequencies
inside the bandwidth of the cavity. This non-classical behavior
exists also without the plate. However, the linear coupling --
even for a plate rotated by a very small angle -- facilitates the
finding of experimental parameters for which entanglement is
observed. When our setup is operated above threshold, frequency
degenerate operation is obtained in a small locking zone and not
only for a precise value of experimental parameters \cite{mason},
which corresponds also to the required parameters below threshold.
When this zone is found, the OPO can be operated below threshold
and the entanglement is maximized by fine tuning of the crystal
temperature. Furthermore, the degenerate operation with bright
beams opens the possibility to match the homodyne detection
without infrared injection of the OPO. In prior experiments with
KTP crystals and green pump, pairing of crystals in order to
compensate walk-off \cite{grangier} or using $\alpha$-cut KTP for
non-critical phase matching and frequency-doubled Nd:YAP laser
\cite{yap} were necessary to generate entangled states of light.

Instead of directly measuring the quantum correlations and
anti-correlations of the signal and idler fields, $A_{1}$ and
$A_{2}$, we characterize the noise of the superposition modes
oriented $\pm45^{\circ}$ from the axes of the crystal
\begin{eqnarray} A_{+}=\frac{A_{1}+A_{2}}{\sqrt{2}} \qquad
\textrm{and} \qquad A_{-}=\frac{A_{1}-A_{2}}{\sqrt{2}} \nonumber
\end{eqnarray}
These two modes have squeezed fluctuations due to the correlations
and anti-correlations between signal and idler fields. The amount
of entanglement between signal and idler can be inferred from the
amount of squeezing available on these superposition modes. We
have developed in \cite{longchambon} a theoretical study of this
original device named "self-phase-locked OPO" in the above
threshold regime. Below threshold, theoretical and experimental
behavior for various birefringent plate angles will be reported on
in a forthcoming paper \cite{laurat04c}.

\begin{figure}[htpb!]
\includegraphics[width=.95\columnwidth,clip=]{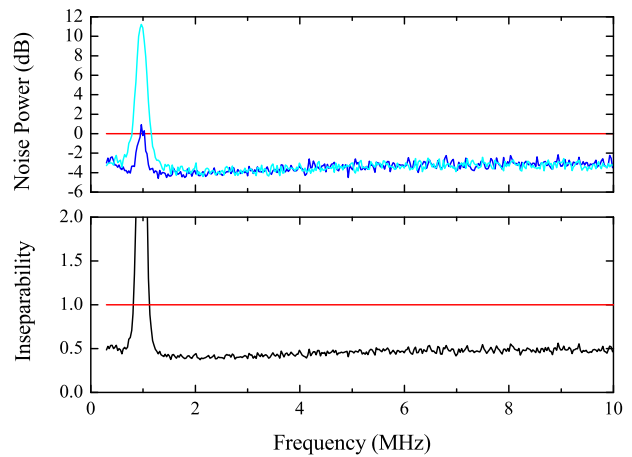}%
\caption{\label{large}Normalized noise variances from 300 kHz to
10 MHz of the $\pm 45^{\circ}$ vacuum modes and inseparability
criterion for signal and idler modes defined as the half-sum of
the previous squeezed variances. The resolution bandwidth is set
to 100 kHz and the video bandwidth to 300 Hz.}
\end{figure}

The experimental setup is shown in Figure \ref{setup}. A
continuous-wave frequency-doubled Nd:YAG laser ("Diabolo" without
option "noise eater", Innolight GmbH) pumps a triply resonant
type-II phase-matched OPO, made of a semi-monolithic linear cavity
: in order to increase the mechanical stability and reduce the
reflection losses, the input flat mirror is directly coated on one
face of the 10 mm-long KTP crystal ($\theta=90^{\circ}$,
$\varphi=23.5^{\circ}$, Raicol Crystals Ltd.). The reflectivities
for the input coupler are 95.5\% for the pump at 532 nm and close
to 100\% for the signal and idler beams at 1064 nm. The output
coupler with a radius of curvature of 38 mm is highly reflecting
for the pump and its transmission is 5\% for the infrared. A
birefringent plate - $\lambda/4$ for the infrared and almost
$\lambda$ for the pump - is inserted inside the cavity. Rotation
as small as $0.01^{\circ}$ can be obtained thanks to a rotating
mount controlled by a piezo-electric actuator. At exact triple
resonance and for a very small angle of the plate relative to the
axes of the crystal, the oscillation threshold is less than 20 mW,
close to the threshold without the plate \cite{laurat03}. The OPO
is actively locked on the pump resonance by the Pound-Drever-Hall
technique: we detect by reflection a 12 MHz modulation and the
error signal is sent to a home-made proportional-integral
controller. The OPO can operate stably during more than one hour
without mode-hopping using a drastic control of the crystal
temperature within the mK range and an optical table isolated from
floor vibrations by pneumatic feet (Newport I-2000).

The 1064 nm laser output is used as a local oscillator for
homodyne detection of the generated state. This beam is spatially
filtered and intensity-noise cleaned by a triangular-ring 45
cm-long cavity with a high finesse of 3000, which is locked on the
maximum of transmission by the tilt-locking technique
\cite{Shaddock}. The homodyne detection is based on a pair of
balanced high quantum efficiency InGaAs photodiodes (Epitaxx
ETX300, quantum efficiency: 95\%) and the fringe visibility
reaches 0.97. The shot noise level of all measurements is easily
obtained by blocking the output of the OPO. A half-wave plate
inserted at the output of the OPO, just before the first
polarizing beam splitter of the homodyne detection, enables us to
choose the field to characterize: signal or idler modes which are
entangled, or the $\pm45^{\circ}$ rotated modes which are
squeezed. The homodyne detection can be locked on the squeezed
quadrature during more than an hour using a standard dither and
lock technique. The error signal is extracted from the demodulated
noise at a given frequency after enveloppe detection.

Figure \ref{large} shows the experimental broadband noise
reduction observed in the $\pm45^{\circ}$ vacuum modes for
frequencies between 300 kHz and 10 MHz. One observes that these
two modes are squeezed well-below the standard quantum limit,
except around 1 MHz where the narrow peak of excess noise is due
to the relaxation oscillation of the laser. One can note that this
excess is less important on the mode $A_{+}$ which is sensitive to
phase noise of the laser. A noise eater implemented on the Nd:YAG
laser should permit to largely reduce this classical excess noise.
The degree of entanglement between signal and idler fields can be
evaluated by the inseparability criterion developed by Duan
\textit{et al.} \cite{duan} and Simon \cite{simon}: a necessary
condition for inseparability is that the half-sum of the previous
squeezed variances falls below one. This criterion is well
verified in the considered frequency band as one can observe in
Figure \ref{large}. To the best of our knowledge, our setup
generates the best EPR entangled beams ever produced in the
continuous variable regime. We have measured at a given noise
frequency of 3.5 MHz (RBW set to 100 kHz and VBW to 300 Hz) a
value of the inseparability criterion equal to $0.33\pm0.02$
\cite{laurat04c}.

\begin{figure}[htpb!]
\includegraphics[width=0.95\columnwidth,clip=]{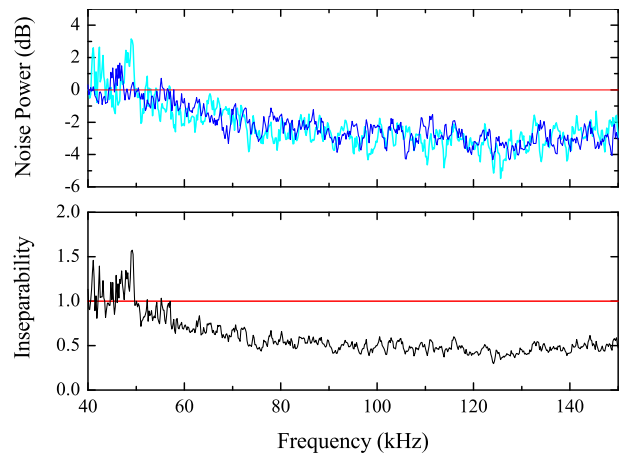}%
\caption{\label{bf}Normalized noise variances from 40 kHz to 150
kHz of the $\pm 45^{\circ}$ vacuum modes after correction of the
electronic noise and inseparability criterion for signal and idler
modes. Squeezing and entanglement are observed down to 50 kHz. The
resolution bandwidth is set to 3 kHz and the video bandwidth to 10
Hz.}
\end{figure}

Figure \ref{bf} sums up the same experimental measurements as
given by Figure \ref{large} but now for low noise frequencies,
between 40 kHz and 150 kHz. Measurements are corrected from the
electronic dark noise which is at least 4 dB below all traces. One
can see that the $\pm 45^{\circ}$ vacuum modes are strongly
squeezed: the noise variances are still reduced by 3 dB around 100
kHz and reach the shot noise limit for frequencies below 50 kHz.
No significant difference is observed between the two rotated
modes. The low limit frequency is well below the ones previously
reported. In contrast to \cite{bowen} and \cite{schnabel}, the low
frequency squeezing is obtained with the same efficiency on both
rotated modes, showing that the effect is not due to common mode
rejection of excess noise but to intrinsic absence of low
frequency noise in our setup.

We have reported in this letter the demonstration of a compact
source of squeezing from 50 kHz to 10 MHz (still present above, in
the bandwidth of the cavity, but not measured) with a slight
increase of noise in a frequency band of 100 kHz around 1 MHz.
This broadband noise reduction is likely to reduce noise in a
pulsed measurement during a time window of duration T. The noise
variance $\sigma^{2}$ of the measurement can be written
\begin{eqnarray}
\sigma^{2}=\int_0^{+\infty} S(\nu) T^{2} sinc^{2}(\pi \nu T) d\nu
\nonumber
\end{eqnarray}
where $S(\nu)$ is the spectral noise density of the light source.
Let us evaluate the improvement obtained on a measurement during a
window of duration $T=1 \mu s$. We model our device by a
shot-noise limited source below 50 kHz (the excess noise can be
reduced to shot-noise by a feed-back loop) and squeezed by 3 dB
above. In comparison with a shot-noise limited source at all
frequencies, the noise variance is divided by a factor 1.7. This
very simple example shows the great interest of broadband and
low-frequency squeezing to improve a large class of physical
measurements.

In summary, we have demonstrated significant broadband vacuum EPR
entanglement and squeezing down to 50 kHz with a single OPO below
threshold and without the need of optical noise cancellation
technique. The degree of entanglement can be improved further by
increasing the transmission escape efficiency of the OPO. Up to
now, the attainable lowest frequency seems only limited by the
technical noise of the laser and locking noise of the different
cavities involved in our setup. The implementation of intensity
noise eater on the Nd:YAG laser - as it is the case for
gravitational wave detectors where lasers are expected to be shot
noise limited around a few tens of Hz - should allow to reach even
lower frequencies.

\begin{acknowledgments}
Laboratoire Kastler-Brossel, of the Ecole Normale Sup\'{e}rieure
and the Universit\'{e} Pierre et Marie Curie, is associated with
the Centre National de la Recherche Scientifique (UMR 8552).
Laboratoire Mat{\'e}riaux et Ph{\'e}nom{\`e}nes Quantiques is a
F{\'e}d{\'e}ration de Recherche (CNRS FR 2437). This work has been
supported by the European Commission project QUICOV
(IST-1999-13071) and ACI Photonique (Minist\`ere de la Recherche).
\end{acknowledgments}

\end{document}